# Efficient Resource Matching in Heterogeneous Grid Using Resource Vector

Srirangam V Addepallil[1], Per Andersen [2] and George L Barnes[3]

[1]High Performance Computing Centre, Texas Tech University, Lubbock, Texas
mailto:srirangam.v.addepalli@ttu.edu

[2]Department of Computer Science, Texas Tech University, Lubbock, Texas
mailto:per.andersen@ttu.edu

[3]Department of Chemistry, University of Oregon, Eugen, Oregon
mailto:gbarnes@uoregon.edu

## ABSTRACT

*In this paper, a method for efficient scheduling to obtain optimum job throughput in a distributed campus grid environment is presented; Traditional job schedulers determine job scheduling using user and job resource attributes. User attributes are related to current usage, historical usage, user priority and project access. Job resource attributes mainly comprise of soft requirements (compilers, libraries) and hard requirements like memory, storage and interconnect. A job scheduler dispatches jobs to a resource if a job's hard and soft requirements are met by a resource. In current scenario during execution of a job, if a resource becomes unavailable, schedulers are presented with limited options, namely re-queuing job or migrating job to a different resource. Both options are expensive in terms of data and compute time. These situations can be avoided, if the often ignored factor, availability time of a resource in a grid environment is considered. We propose resource rank approach, in which jobs are dispatched to a resource which has the highest rank among all resources that match the job's requirement. The results show that our approach can increase throughput of many serial / monolithic jobs.*

.

## 1. INTRODUCTION

A grid computing environment exhibits three broad characteristics.

1. Resource Co-ordination

2. Resource integration

3. Nontrivial QoS [10]

In a Grid computing environment all computation resources are managed by resource managers such as PBS (Portable Batch Scheduler) [15] [16], Load Sharing Facility (LSF) [14], Sun Grid Engine [18] and Load Leveller [17]. These are local resource managers that interface with a Global scheduler, referred to as a Metascheduler. A metascheduler identifies resources and dispatches a job to candidate resource. Current metaschedulers are unable to overcome resource availability factor in grid environment. Thus, new scheduling paths/mechanisms are needed that will take into account resource availability; and schedule user jobs in an efficient manner.

Applications dispatched to a grid environment have their own resource requirements; these requirements are not uniform for all applications. Hence the performance and throughput of the application differs greatly from application to application (and the manner in which resources are allocated to these jobs). Data input file values and number of input parameters that are being provided to the same type of application specific job also have a significant effect on the job completion time. In grid computing the wide variety of resources in terms of machines, storage, interconnects, software licensing and libraries make the job matching task even more complicated. Clearly there is a need for efficient resource matching that takes into account a large variety of resources available. In this paper resource vector based approach is presented that aims at identifying the best resource for a job dispatched.

The metascheduler dispatches jobs to resource using a Grid Resource Vector (GRV) for matching resource. The GRV is updated constantly depending on the resource usage, work done, and length of availability. We evaluate this mechanism through extensive simulation using real work traces, real time user job submission and observe how the Meta scheduler manages several jobs.

## 2. RELATED WORK

Grid Computing has evolved from individual systems to clusters and pool of systems/clusters using several techniques including cycle scavenging to provision resources all working in conjunction that are transparent to end user. Resource reliability, Job completion, and local scheduling are some external facts that have been overlooked in classical scheduling approaches. These factors pay a major role in job completion and directly effect through put. Moab [20] an advanced Meta scheduler that allows distributed work load to be run across independent clusters. Grid way [21] is a light-weight meta-scheduler that follows greedy approach to schedule user jobs in a First come first server (FIFO) manner. Condor and Condor-G [23] are specialized workload management systems for compute-intensive jobs. How ever none of the above scheduling mechanisms support scheduling based on a QoS/Feedback/ Reliability mechanism. gLite[24] workload management system is used to manage large computing installation. gLite uses eager and lazy scheduling policies for scheduling jobs to individual resources for execution. In eager scheduling a job is bound to a resource as soon as possible and the decision has taken. Lazy scheduling waits for a resource to become available before it is matched to a submitted job. There are other schedulers like Community scheduler frame work which co-ordinates communications among multiple heterogeneous schedulers that operate at cluster level. It interfaces with LSF, open PBS and Sun Grid Engine

Many resource allocation approaches have been proposed for Distributed systems and Grids. Auction based resource allocation uses a bid price for user jobs and allocates resource to the highest bidder. [1]. Ontology-based resource matching techniques [3] use a rank based approach, where rank is obtained by the hierarchical access to the resource by users. Here a resource belonging to a priority group will get a higher rank than a normal resource, local users get higher priory over remote users. Agent based resource allocation [2] uses agents to identify appropriate resources. Other user centric techniques like resource allocation Hierarchy [4].In all the above approaches jobs are matched to resources according to requirements and attributes. Availability (Reliability) of the resource has always been ignored. Unlike earlier approaches where jobs are matched to resource, our approach matches both jobs to resources and resource to jobs. Our work is different from other resource matching mechanisms as the GRV is transparent to enduser and resources. This mechanism computes GRV for resources based on dynamic conditions such as number of success full job completion and continuous availability.

## 3. SCHEDULING SCHEME

In this section we present a metascheduler that is typically found in educational institutions. In this model, we have cycle scavenging resources, and dedicated compute nodes. In this model each resource is manager by different administration entities or departments. Resource manger (Condor) is organized as a set of resources that can be accessed using a job submission queue. Access to these resources is controlled by predetermined attributes, in addition to these; cycle scavenging systems have additional flags about job migration when an interactive session is initiated by a user. Condor, SGE and LSF provide configurable options (preempt-evict, checkpoint-migrate) to address these situations. Discussing all these interactions will be out of scope for this paper.

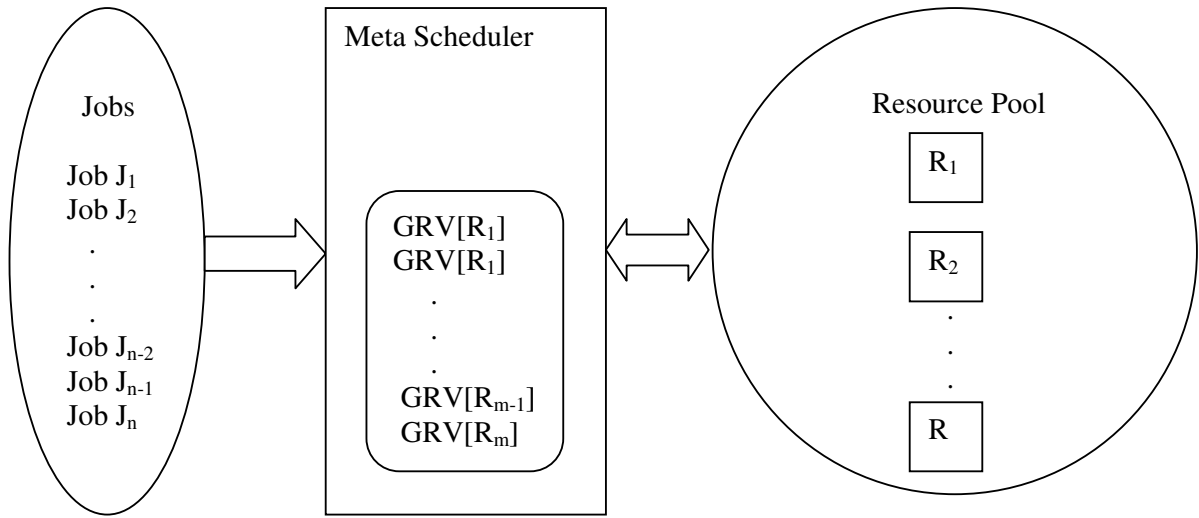

**Figure 1: Schema of Metascheduler using Grid Resource Vector**

Metascheduler maintains a record for each resource in the compute grid GRV. All resources start with same initial priority and are listed as possible candidates for job execution. When users submit jobs to Metascheduler along with dependencies for execution, Metascheduler identifies a set of candidate machines satisfying resource requirements. Resource with the highest GRV value is identified as a candidate machine and jobs is dispatched to candidate machine. In traditional closely integrated cluster reliability and job success ratio do not play a significant role in resource selection, but in a grid environment where resources are typically available for a few hours or when dynamic provisioning of CPU cycles is in effect they become the deciding factor.In the current system we assume that a user jobs are independent of each other and can be run on grid resources. Information is communicated back to Meta scheduler, when the job is completed, evicted or exit. Metascheduler, update's GRV of specific resources based on the job's exit status. As new jobs come in the priority of resource change. Figure 1 shows the mechanism for calculating and maintaining GRV on Meta scheduler.

## 4. GENERATING GRID RESOURCE VECTOR: GRV

A GRV employs three main attributes: Resource Availability ($RA_m$), Job success ($JS_m$) and custom attribute ($CA_m$) for each resource. These three attributes are combined together can be used to implement several resource matching paths and a range of scheduling concepts. After an initial configuration, the Metascheduler controls resource matching using GRV without further administrative intervention.

$GRV[m] = RA_m + JS_m + CA_m$

The three attributes and our approach for combining these three attributes to derive a final resource priority are described in this section. Resource Availability ($RA_m$), of a resource is determined based on the length of time it is continuously available for job execution. Here a resource that has just showed up on compute grid will just have base priority, which is lower than a resource that has been available previously, A resource that has been available over a period of time will have priority that is less than or equal to a defined maximum resource priority. To avoid assigning infinite priority to a resource that has been available for a long time, we use the above approach.

Job success ($JS_m$) accounts for the application execution part of the job. A successful completion of job $J_1$ on resource M will result in incrementing $JS_m$ of attribute of resource M, any other situation like evict, requeue or checkpoint operations result in a Penalty and $JS_m$ is decremented. Hence resources that have successfully completed execution many jobs have a higher ($JS_m$).

Custom Attribute ($CA_m$) is used to match a job to the resource with higher system performance. ($CA_m$) of machine m can be determined using CPU clock speed, no of cores. These two attributes gives us the number of floating point operations that can be performed on a resource. ($CA_m$) can be fine tuned to include other resource attributes Memory, SWAP, Storage. It will again be out of scope of this paper to enumerate on these. For simplicity we consider ($CA_m$) to be influenced only by floating point operations.

$CA_m = nflops_m * nc_m$

Where $nflops_m$ is the number of floating point operations that can performed on machine m and $nc_m$ is number of cores on the resource. In closely coupled clusters, all resources are homogenous, so ($CA_m$) does not pay a huge factor. In a grid, because of heterogeneous system Custom Attribute ($CA_m$) carries significant weight. Hence depending on scenario the three attributes Resource Availability ($RA_m$), Job success ($JS_m$) and Custom Attribute ($CA_m$) have varying degrees of influence.

To address this issue we assign a weight to each attribute. Equation 4 will take the form $GRV[M] = W_1*RA_m + W_2 * JS_m + W_3 * CA_m$ Choice of weight factors determines the final value of GRV[M] and indentifying candidate machine.

**Implementation Algorithm 1: Code for GRV**

1. *Initialize $GRV[M_n]=0$*
2. *Obtainresource(Scan )*
3. *If initial-setup; /* For initial setup of vector */*
4. *Then*
5. *While m from 1,n ;do*
6. *$CA_i = nflops_m * nc_m$*
7. *$JS_m$ = intialval*
8. *$RA_m$ = rbase*
9. *end while*
10. *else   /* update resource vector */*
11. *while m from 1, n ;do*
12. *update ($GRV[M_l]$)*
13. *end while*

*14  end if*

*15  List jobs ← All_queued_jobs.*
*16  For j in list jobs do*

*17       PossibleCandidates ← AllUNoccupiedMatchingResources[M]*

*18  For p in PossibleCandidates*

*19  CandidiateMachine ← Max [GRV[PossibeCandidates]]*

*20  JobDispatch (Candidate Machine)*

*21  End for*

*22  End for*

*23  JobDispatch(Candidate machine)*

*24  If Complete then*

*25  Update($JS_M=JS_M+Reward$)*

*26  Else*

*27  Update($JS_M=JS_M-Penanlty$)*

*28  End if*

## 5. EXPERIMENTAL SETUP

We use condor pools to implement our Grid resource vector based mechanism, for matching user jobs with Grid resources. Our implementations use observations gathered in real time. For this setup two computational chemistry applications CSTechG [11] and VENUS [12] were identified. 17300 individual jobs form different users have been recorded that requested VENUS or CSTechG, and were dispatched to resources. Run time, exit status and host machine attributes re recorded for each user job j.

Grid computing infrastructure comprised of 546 heterogeneous distributed resources spread across 8 different departments, connected via network. Resources were broadly classified into three categories. Following is a snap shot of resource in real time which shows matched resources and claimed resources.

Table 1 Campus grid resources

| TOTAL | Owner | Claimed | Unclaimed | Matched |
|---|---|---|---|---|
| INTEL/LINUX | 4 | 0 | 0 | 4 |
| INTEL/WINNT5 | 538 | 97 | 0 | 441 |
| X86_64/LINUX | 4 | 0 | 0 | 4 |
| Total | 546 | 97 | 0 | 449 |

For the purpose of this test case, it is assumed that resource in certain classification have similar performance Custom Attribute ($CA_m$) and does not vary hugely in between resources. Hence the GRV priority between different machines is directly proportional to the weights assigned to Resource Availability ($RA_m$), and Job success ($JS_m$). In the following test cases equal weights have been assigned to W1 and W2. Dedicated compute cluster It has total of 2312 CPUs, 210 dual-slot quad-core nodes with Intel E5405 processors for a total of 1680 3.0 GHz cores, connected with DDR Infiniband, access is controlled using queues and job run time with in queues. Each queue had a varying run length; certain queues let jobs runs for up to 48 hours, some for 120 hours. Upon reaching the runtime limit jobs are check pointed terminated and returned to top of the wait queue. Queues with lower run time limit have higher priority and queues with higher run time limit have lower priority. If a job with lower run time requirements is submitted to lower priority queue the time it waits in the queue determines the total turn around time. Hence queue selection is important in the case of homogeneous closely coupled clusters. We compare the results of jobs submitted in a dedicated compute cluster and in distributed heterogeneous grid system using GRV and FCFS approaches.

**Application 1: CSTechG**
In General simulation of one single chemical reaction requires myriads of trajectories in the CS quantum phase space [11]. These dynamics can be naturally implemented for simultaneous trajectory runs on a compute grid. Current application CSTechG developed for several operating systems (Microsoft Windows®, Red Hat Linux® ) from the ENDyne 2.7 and 2.8 codes [13]. From deployment point of view, a CSTechG is capable of performing the following tasks: (1) define the reaction initial conditions from any grid node, (2) submit all the t trajectory jobs from those initial conditions and run all trajectories in parallel on any of the available grid nodes; All the monolithic jobs run in parallel and contribute to final outcome of the runs. Hence the completion time of the slowest job determines the final outcome of the set of jobs; hence effective resource matching for this job(s) will ensure higher levels of through put. Here Throughput is the percentage of total number of job submissions to percentage of jobs completed successfully without resubmission In this test case we assign equal weights to all three attributes $W_1, W_2, W_3 = 0.33$

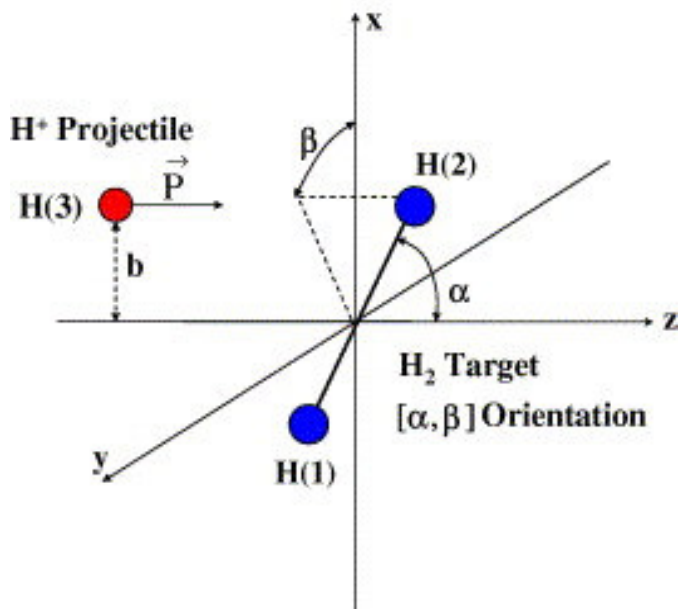

**Figure 2: $H^+ + H_2$ reactants initial conditions.**
Balls represent nuclear wave packets for the atoms with projectile impact parameter b and target orientation [a, b]. [19]

Table 2 CSTechG job submission information

|  | Case 1 | | | | Case 2 | | | |
|---|---|---|---|---|---|---|---|---|
| Job Type | Total Jobs | Jobs Completed | Restarted | Throughput | Total Jobs | Jobs Completed | Restarted | Throughput |
| GRV | 6004 | 5491 | 509 | 91.5 | 1010 | 769 | 231 | 70 |
| NonGRV | 6004 | 4917 | 512 | 81.9 | 1010 | 590 | 420 | 59.0 |
| Cluster | 6004 | 5845 | 159 | 97.3 | 1010 | 1001 | 9 | 99.9 |

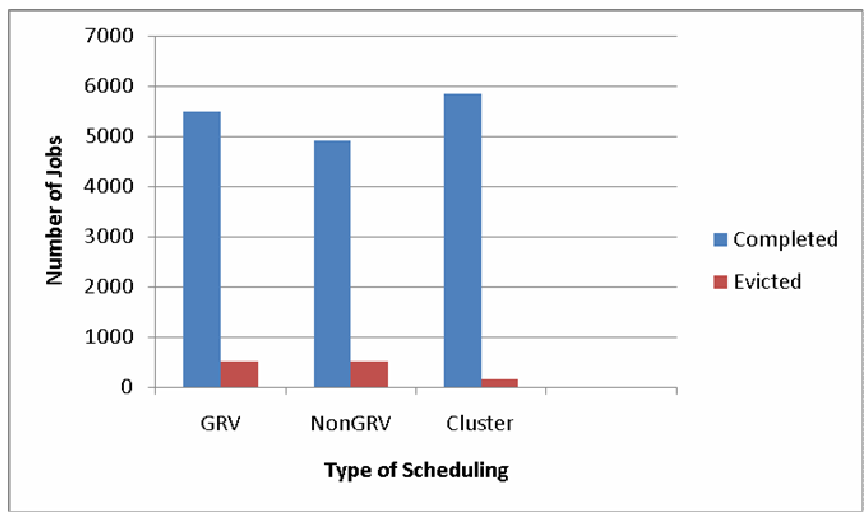

**Figure 3: CSTechG using GRV,NonGRVand Cluster Scheduling Case 1**

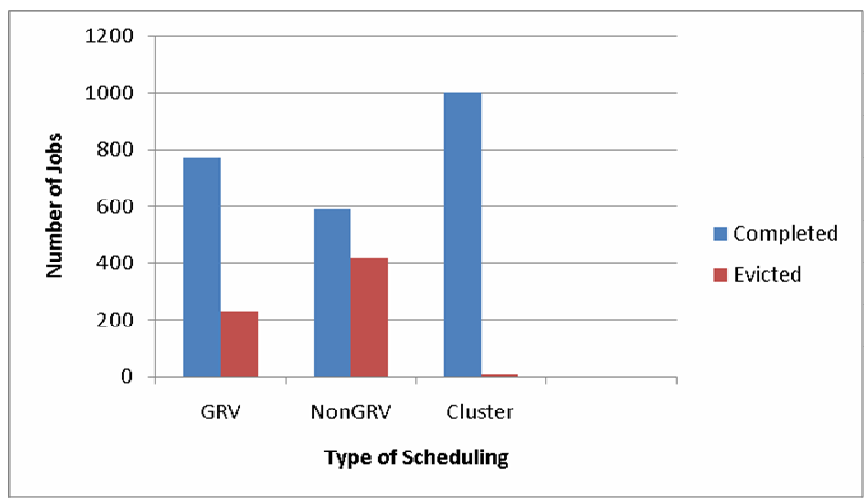

**Figure 4 CSTechG using GRV,NonGRVand Cluster Scheduling Case 2**

From the above set's of run's it is observed that using GRV for resource matching, helps us achieve higher throughput than conventional resource matching. Throughput for dedicated cluster is closer to 100. This emphasizes that resource reliability is also a major factor in determining through put.

**Application 2: VENUS/VENUS-MOPAC**
At a zero order level much of chemistry and chemical reactivity can be described through use of classical mechanics if an accurate potential energy surface (PES) is used in the propagation of Newton's equations of motion. This is actually a big IF as the dimensionality of a PES scales as ~3N where N is the number of atoms. If only 2 points (an absurdly small number, 10 or 20 is more reasonable) are required to accurately evaluate and fit the PES then $2^{3N}$ calculations would be required. For system sizes much larger that 5 or 6 atoms it becomes computationally prohibitive to evaluate the PES in advance. ``Direct Dynamics" solves this problem by calculating the necessary PES information during the propagation of Newton's equations. Venus Mopac consists of a well developed molecular dynamics code (Venus), which allows for many different methods of selection of initial conditions as well as various integrators, coupled to a semi-empirical quantum mechanical package (Mopac) which provides PES information for a given configuration of atoms. The computational time required mostly depends on the evaluation of the PES, though the number of integration cycles and the integration method are also factors. Trajectories with the same integration method and number of cycles will take roughly the same amount of time to calculate, though differences are possible since some atomic configurations require more computational expense to obtain the PES.

Table 2 CSTechG job submission information

| Type | Total Jobs | Jobs Completed | Restarted | Throughput | Total Jobs | Jobs Completed | Restarted | Throughput |
|---|---|---|---|---|---|---|---|---|
| GRV | 1237 | 1073 | 164 | 86.7 | 5873 | 3969 | 1904 | 67.58 |
| NonGRV | 1237 | 883 | 354 | 71.38 | 5873 | 4790 | 1083 | 81.55 |
| Cluster | 1237 | 1221 | 16 | 99.9 | 5873 | 5631 | 242 | 96.5 |

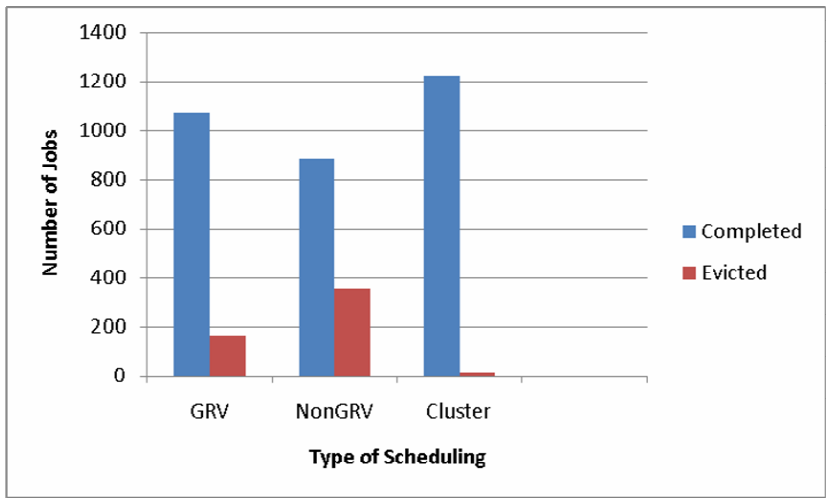

**Figure 5 VENUS using GRV,NonGRVand Cluster Scheduling  Case 1**

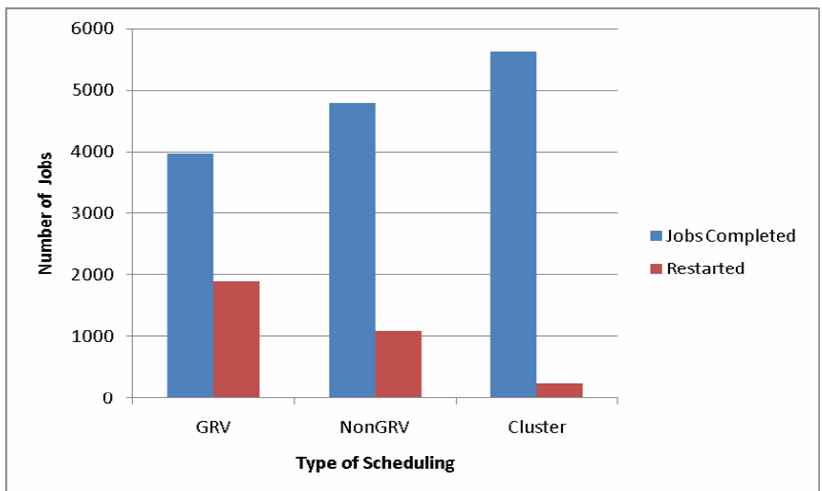

**Figure 6 VENUS using GRV,NonGRVand Cluster Scheduling  Case 2**

## 6 Conclusions

In this paper a Grid Resource Vector based approach has been presented. The meta scheduler uses GRV with different weights to match resource to user applications in an efficient manner to increase over all through put. The GRV is dynamic in nature and changes according to resource reliability and job completion. We compare our approach to traditional approach FCFS that is employed currently. Our results clearly show that our mechanism will result in increased throughput for monolithic/serial applications and indirectly increasing throughput of other user applications. When compared to results to a homogeneous compute cluster (e.g.: A rocks compute cluster), where through put is in higher 90's GRV approach gives better results than non GRV approach.

It is highly unlikely that we might achieve throughput of the order achieved in a dedicated homogenous compute cluster, how ever in future we would like to develop a mechanism in obtaining a approximate time / task execution and use in conjunction with GRV to more efficiently match jobs to resource and increase throughput further.

## ACKNOWLEDGMENTS
The Authors would like to thank Jerry Perez, Grid Administrator at high performance computing centre for his assistance in setting up the testing system, Dr. Buddha Maiti for using job submission log information.